\documentclass[aps,prc,twocolumn]{revtex4}

\include{setspace}
\include{dcolumn}
\usepackage{graphicx}
\usepackage{longtable}
\usepackage{multirow}
\usepackage{array}
\begin{document}

\title{Giant Monopole Resonance in even-A Cd isotopes, the asymmetry term in nuclear incompressibility, and the ``softness'' of Sn and Cd nuclei}

\author{D.~Patel$^{1}$,\, U.~Garg$^{1}$,\, M.~Fujiwara$^{2}$,\, H.~Akimune$^{3}$,\, G.P.A.~Berg$^{1}$,\, M.N.~Harakeh$^{4}$,\, M.~Itoh$^{5}$,\, T.~Kawabata$^{6}$,\, K.~Kawase$^{7}$,\,  B.K.~Nayak$^{1}$\footnote{Permanent address: Nuclear Physics Division, Bhabha Atomic Research Center, Mumbai 400 085, India.},\, T.~Ohta$^{2}$,\, H.~Ouchi$^{5}$,\, J.~Piekarewicz$^{8}$,\, M.~Uchida$^{9}$,\, H.P.~Yoshida$^{5}$\, and M.~Yosoi$^{2}$}

\affiliation{$^1$Department of Physics, University of Notre Dame, Notre Dame, Indiana 46556, USA \\$^{2}$Research Center for Nuclear Physics, Osaka University, Osaka 567-0047, Japan \\$^{3}$Department of Physics, Konan University, Kobe 568-8501, Japan \\$^{4}$Kernfysisch Versneller Instituut, University of Groningen, 9747 AA Groningen, The Netherlands \\$^{5}$Cyclotron and Radioisotope Center, Tohoku University, Sendai 980-8578, Japan \\$^{6}$Center for Nuclear Studies, University of Tokyo, Tokyo 113-0033, Japan \\$^{7}$Institute of Scientific and Industrial Research, Osaka University, Osaka 567-0047, Japan \\$^{8}$Department of Physics, Florida State University, Tallahassee, FL 32306, USA \\$^{9}$Department of Physics, Tokyo Institute of Technology, Tokyo 152-8850, Japan}

\date{\today}

\begin{abstract}

The isoscalar giant monopole resonance (ISGMR) in even-A Cd isotopes has been studied by inelastic ${\alpha}$-scattering at 100 MeV/u and at extremely forward
angles, including $0^\circ$.
The asymmetry term in the nuclear incompressibility extracted from the ISGMR in Cd isotopes is found to be $K_{\tau} = -555 \pm$75 MeV, confirming the value
previously obtained from the Sn isotopes. ISGMR strength has been computed in relativistic RPA using NL3 and FSUGold effective interactions. Both models significantly overestimate the centroids of the ISGMR strength in the Cd isotopes.
Combined with other recent theoretical effort, the question of the ``softness'' of the open-shell nuclei in the tin region remains open still.
\end{abstract}

\maketitle

The equation of state (EOS) of nuclear matter plays an important role in our understanding of a number of interesting phenomena such as the collective behavior of nucleons in the nuclei, the massive stellar collapse leading to a supernova explosion, nuclear properties including the neutron-skin thickness of heavy nuclei, and the radii of neutron stars \cite{Prakash2001,HoroPiekare2001}. The nuclear incompressibility, $K_\infty$, is the curvature of EOS of nuclear matter at saturation density \cite{BohrMott1975}. $K_{\infty}$ is, thus, a measure of nuclear stiffness and thereby imposes significant constraints on theoretical descriptions of the effective nuclear interactions. However, even more stringent constraints
emerge as one studies the evolution of the incompressibility coefficient as the system becomes neutron rich. Neutron-rich systems are sensitive to the poorly-known density dependence of the symmetry energy and the experiments reported here are of vital importance in this regard.

The study of the isoscalar giant monopole resonance (ISGMR) provides a direct experimental tool to study nuclear incompressibility in finite nuclear systems. The centroid energy of ISGMR, $E_{ISGMR}$, can be directly related to the nuclear incompressibility of finite nuclear matter, $K_A$, as:
\begin{equation}\label{eq:2}E_{ISGMR}=\hbar\sqrt{\frac{K_A}{m<r^2>}}
\end{equation}
\noindent
where, $m$ is the nucleon mass and $<r^2>$ is the mean square radius of the nucleus \cite{Strin82,trein81}.
$K_A$ may be further parameterized as \cite{Blazoit80,trein81}:
\begin{equation}\label{eq:3}{K_A\simeq K_{vol}(1+cA^{-1/3})+K_\tau((N-Z)/A)^2+K_{Coul}A^{-4/3}}
\end{equation}
\noindent
Here, $K_{vol}$ is the volume term, directly related to $K_{\infty}$, $c\sim -1$ \cite{Patra2002}, $K_{Coul}$ is essentially a model-independent term \cite{Sagawa2007}, and $K_{\tau}$ is the asymmetry term.
Although closely related, the finite-nucleus asymmetry term $K_{\tau}$ should not be confused with the corresponding term in infinite nuclear matter--a quantity also denoted by $K_{\tau}$ at times, but written here as
$K_{\tau}^{\infty}$. Indeed, $K_{\tau}^{\infty}$ should never be regarded as the $A\!\rightarrow\!\infty$ limit of the finite-nucleus asymmetry $K_{\tau}$ \cite{Blazoit80}. Yet the fact that $K_{\tau}$ is both experimentally accessible and strongly correlated with $K_{\tau}^{\infty}$ is vital in placing stringent constraints on the
density dependence of the symmetry energy. Recall that $K_{\tau}^{\infty}$ is simply related to a few fundamental parameters of the equation of state \cite{cente09}:
\begin{equation}
 K_{\tau}^{\infty}= K_{\rm sym}-6L-\frac{Q_{0}}{K_{\infty}}L\,,
\end{equation}
where $Q_{0}$  the {\sl ``skewness''} parameter of symmetric nuclear matter and $L$ and $K_{\rm sym}$,
respectively, are the slope and curvature of the symmetry energy. It is the strong sensitivity of $K_{\tau}^{\infty}$ to the density dependence of the symmetry energy that makes the present study of critical importance in constraining the EOS of neutron-rich matter.

This asymmetry term,  $K_{\tau}$, can be studied over a series of isotopes for which the neutron-proton asymmetry, $(N-Z)/A$, changes by an appreciable amount. The first such investigation was carried out by
Sharma et al. \cite{sharma} and later much improved by Li et al. over the even-$A$ $^{112-124}$Sn isotopes \cite{Tao2007,Tao2010}. Li et al. obtained a value
$K_{\tau} = -550 \pm$100 MeV, a number consistent with values indirectly obtained from several other measurements
\cite{Sagawa2007,Chen2009,Centelles2009}. Most intriguingly, the ISGMR centroid energies of the Sn isotopes were found to be consistently lower than those predicted by relativistic and non-relativistic calculations by as much as 1 MeV \cite{Tao2007,Tao2010}. It bears noting that the interactions used in these calculations were the very same that reproduced the ISGMR centroid energies in ``standard'' nuclei, $^{208}$Pb and $^{90}$Zr, very well, leading to the question: {\sl Why are the Sn isotopes so fluffy?} \cite{Piekare2007, garg2007}.

To confirm the value of $K_{\tau}$ obtained from the Sn isotopes, and to further explore the lack of success of the relativistic as well as non-relativistic calculations in correctly obtaining the centroids of the ISGMR strength in the open-shell nuclei, we have measured the ISGMR strength distributions in the even-$A$ $^{106,110-116}$Cd isotopes. The asymmetry parameter, $(N - Z)/A$, changes by as much as $83\%$ along this chain, making these nuclei very attractive, like the stable Sn isotopes, from the point of view of extracting the asymmetry term in the nuclear incompressibility. Preliminary results from this investigation have been presented previously \cite{prelim1,prelim2}.

The experiment was performed at the Ring Cyclotron facility of the Research Center for Nuclear Physics (RCNP), Osaka University, Japan. $^4$He particles at an incident energy of 100 MeV/u were scattered off self-supporting even-$A$ $^{106,110-116}$Cd targets; highly-enriched Cd targets ($>$93$\%$) with thicknesses ranging from 5 to 6.5 mg/cm$^2$ were used.
Elastic as well as inelastic scattering measurements were performed over a wide range of angular
settings--elastic scattering over $3.4^\circ-19^\circ$ and inelastic scattering at extremely forward angles ($0^\circ$--$9.8^\circ$). The primary justification for the difficult-to-do extremely forward angle measurements lies in the angular distribution patterns, which exhibit clear distinction between various multipoles at these angles and, with the ISGMR cross sections peaking at $0^\circ$,  it is extremely important to make a measurement as close to $0^\circ$ as possible.

The scattered $\alpha$ particles were momentum analyzed by the high-resolution magnetic spectrometer, Grand Raiden \cite{Fujiwara99}, and focused onto the focal-plane detector system comprised of two MWDCs \cite{Fujita99} and two plastic scintillator counters. The MWDCs allow measurement of the horizontal and vertical coordinates of the impact position of the $\alpha$ particle on the focal plane. This, in turn, allows the determination of the angle of incidence on the focal plane and the momentum of the scattered $\alpha$ particles. Using the ray-tracing technique for trajectory determination of scattered particles, energy spectra were obtained for specific scattering angles by subdividing the full angular opening. The focal-plane detector covered the excitation-energy range of $E_x$ $\sim$ 8 MeV to 31 MeV. Energy calibration runs were carried out with a $^{12}$C target at every angle for each target. The Grand Raiden spectrometer was used in the double-focusing mode in order to eliminate the instrumental background \cite{Itoh2003,Tao2010}.
The background-subtracted spectra at an ``average'' angle of $0.7^\circ$ for various Cd isotopes are shown in Fig.~1.

\begin{figure}
\includegraphics[width=0.48\textwidth]{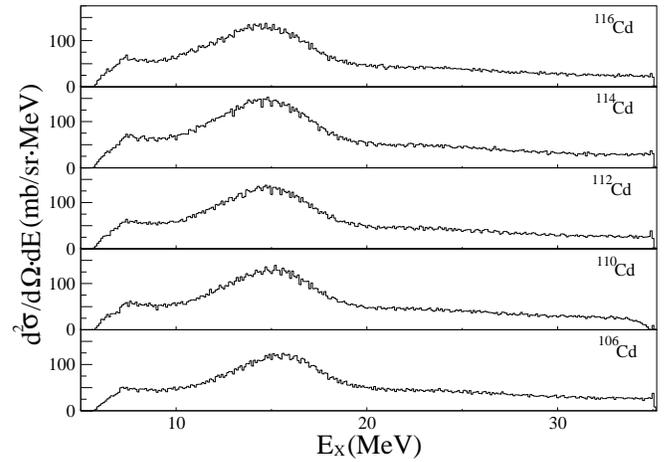}
\caption{Excitation-energy spectra at an ``average'' angle of 0.7$^\circ$ for the even-even Cd isotopes investigated in this work.\label{bc}}
\end{figure}

Elastic-scattering cross sections were used to extract the optical model parameters (OMP) for these beam-target combinations. The parameters were determined using the ``hybrid'' potential proposed by Satchler and Khoa \cite{xxxx} and found to work very well for $\alpha$-scattering at medium energies (see, for example,
Refs. \cite{Tao2010,Nayak2006,Youngblood}). In this procedure, the density-dependent single-folding model with a Gaussian $\alpha$-nucleon potential was used to determine the real part of the optical potential. The
computer codes SDOLFIN and DOLFIN \cite{RickDOLF} were used to calculate the shape of the real part of the potential and the form factors, respectively. For the imaginary term, a Woods-Saxon potential was used and its parameters, together
with the depth of the real part, V, were obtained by fitting the elastic-scattering cross sections using the minimization of chi-square technique, with the help of the computer code
PTOLEMY \cite{Brown80}. Using the known B(E2) values from the literature \cite{nndc} and the OMP thus obtained, the angular distributions of differential cross sections were calculated for the $2^+_1$ states. An excellent agreement between the calculated and experimental
angular distributions of differential cross sections for the $2^+_1$ states established the appropriateness of the OMP.

The inelastic-scattering cross sections were sorted into 1 MeV bins to reduce statistical fluctuations. The experimentally-obtained spectra consist of contributions from various multipoles. In order to extract the
ISGMR contribution from these spectra at different scattering angles, multipole-decomposition analysis (MDA) was performed. The experimental double-differential cross sections are expressed as linear combinations of calculated distorted-wave Born approximation (DWBA) double-differential cross sections for different multipoles as follows:
\begin{equation}\label{eq:4}\frac{d^2\sigma^{exp}(\theta_{c.m.},E_x)}{d\Omega dE}=\sum_{L=0}^{7}a_{L}(E_x)\frac{d^2\sigma^{DWBA}_L(\theta_{c.m.},E_x)}{d\Omega dE}
\end{equation}
\noindent
where $L$ is the order of the multipole and $a_{L}(E_x)$ is the
\begin{table*}[htb]
 \caption{\label{hn} Lorentzian-fit parameters for the ISGMR strength distributions in the Cd isotopes investigated
 in this work. Also presented are the various moment ratios calculated over the excitation-energy range 10.5 MeV--20.5 MeV, as well as those from calculations for the FSUGold, NL3 and SLy5 (without pairing) interactions; $m_k$ is the $k$th moment of the strength
distribution: $m_k = \int \! E^k_x\ \! S(E_x)\, \mathrm{d} x$. For comparison,
ISGMR parameters from Lui et al. (Gaussian fits) are also provided, where
available \cite{Youngblood}.}
\begin{tabular}{c c c c c c|c|c|c|c}
 \hline
 \hline
 Target & E$_{ISGMR}$ (MeV) & $\Gamma$ (MeV) & E$_{ISGMR}$ (MeV)& $\sqrt{m_1/m_{-1}} (MeV) $ & $\sqrt{m_3/m_1}(MeV) $ &\multicolumn{4}{c}{$m_1/m_0$
 (MeV)} \\

   & & & Ref. \cite{Youngblood} & & & Experiment & FSUGold & NL3 & SLy5\footnotemark[1] \\
   \hline
  $^{106}$Cd & 16.50$\pm$0.19 & 6.14$\pm$0.37 & - & 16.06$\pm$0.05 & 16.83$\pm$0.09 & 16.27$\pm$0.09 & 16.73 & 17.25 & 16.92\\
 $^{108}$Cd & - & - & - & - & - & - & 16.65 & 17.17 & -\\
  $^{110}$Cd & 16.09$\pm$0.15 & 5.72$\pm$0.45 & 15.71 $^{+0.11}_{-0.11}$ & 15.72$\pm$0.05 & 16.53$\pm$0.08& 15.94$\pm$0.07 & 16.59 & 17.09 & 16.65\\

  $^{112}$Cd & 15.72$\pm$0.10 & 5.85$\pm$0.18 & - & 15.59$\pm$0.05 & 16.38$\pm$0.06 & 15.80$\pm$0.05 & 16.50 & 17.00 & 16.50\\

  $^{114}$Cd & 15.59$\pm$0.20 & 6.41$\pm$0.64 & - & 15.37$\pm$0.08 & 16.27$\pm$0.09 & 15.61$\pm$0.08 & 16.38 & 16.90 & 16.47\\

  $^{116}$Cd & 15.43$\pm$0.12 & 6.51$\pm$0.40 & 15.17 $^{+0.12}_{-0.11}$ & 15.19$\pm$0.06 & 16.14$\pm$0.07 & 15.44$\pm$0.06 & 16.27 & 16.77 & 16.36\\
  \hline
\end{tabular}
\footnotetext[1] {Ref. \cite{colo2}}
\end{table*}

\noindent
percentage of the energy-weighted sum rule (EWSR) for
multipolarity $L$. DWBA cross sections corresponding to $100\%$ EWSR were calculated using
transition densities and sum rules provided in Refs. \cite{HarakehBook,Satchler87}. DWBA calculations
were performed for up to a maximum angular-momentum transfer of $L$=7; addition of higher angular-momentum-transfer terms resulted in minimal to no change in the strength distributions. Further details on the MDA can be found in Refs. \cite{Bonin84,Uchida2004,Itoh2003}. The isovector giant dipole resonance (IVGDR) contribution was subtracted out of the experimental spectra prior to the fitting procedure. Photonuclear data were used in conjunction with DWBA calculations based on the Goldhaber-Teller model to estimate the IVGDR differential cross section as a function of scattering angle \cite{Data}.

\begin{figure}
\includegraphics[width=0.48\textwidth]{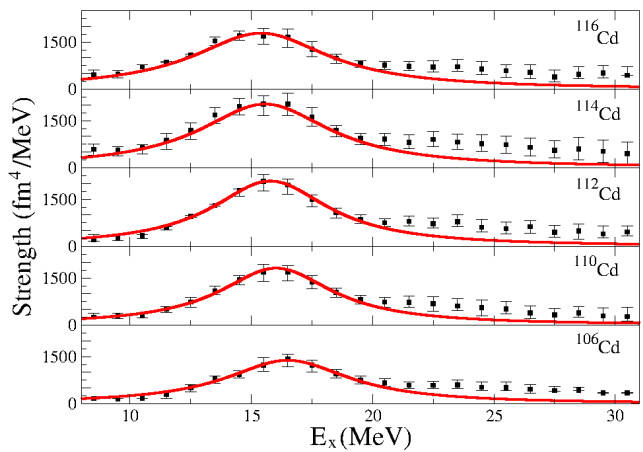}
\caption{(Color online) The ISGMR strength distributions in the Cd
isotopes investigated in this work. The solid lines represent Lorentzian fits to the data. \label{df}}
\end{figure}

The coefficients, $a_{0}(E_x)$, extracted from the MDA were used to obtain the ISGMR strength distributions; these are shown in Fig. 2, along with Lorentzian fits to the data. The choice of Lorentzian shape is arbitrary; the final results are not affected in any significant way if, for example, a Gaussian fit is used instead. Also, as discussed in Refs.~\cite{Tao2007,Tao2010}, the excess strength at the higher excitation energies is attributable to the mimicking of the $L=0$ angular distribution by components of the nuclear continuum from the three-body channels, such as the forward-peaked knock-out process wherein protons and neutrons are knocked out by the incoming $\alpha$ projectiles \cite{bran}. This conjecture is supported by measurements of proton decay from the isoscalar dipole resonance (ISGDR) at backward angles: no such spurious strength was observed in spectra in coincidence with the decay protons \cite{matyas1,Nayak2009}. The parameters of the Lorentzian fit, and the customary moment
ratios, $\sqrt{m_1/m_{-1}}$, $\sqrt{m_3/m_{1}}$, and $m_1/m_0$, computed for consistency from both the experimental and theoretical strength distributions over the energy range 10.5 MeV--20.5 MeV, are provided in Table I. The upper limit of this range has been chosen to minimize contributions from the excess strength starting at around that energy. It should be clear, however, that the comparison between theory and experiment throughout this work is for the same range of excitation energy; this identical energy range is crucial and decisive in the discussion below.

The moment ratios $m_1/m_0$ for the Cd isotopes are presented in Fig.~3. Also displayed are theoretical results extracted from the distributions of isoscalar monopole strength computed in a relativistic random phase approximation (RPA) using the accurately calibrated NL3 ($K_\infty$=271 MeV)\,\cite{Lalazissis:1996rd} and FSUGold ($K_\infty$=230 MeV)\,\cite{Todd-Rutel:2005fa} effective interactions. A detailed description of the relativistic RPA formalism and its implementation may be found in Ref.~\cite{Piekarewicz:2001nm}. We note that the FSUGold model employed here has been successful in reproducing ISGMR

\begin{figure}[h]
\includegraphics[width=0.48\textwidth]{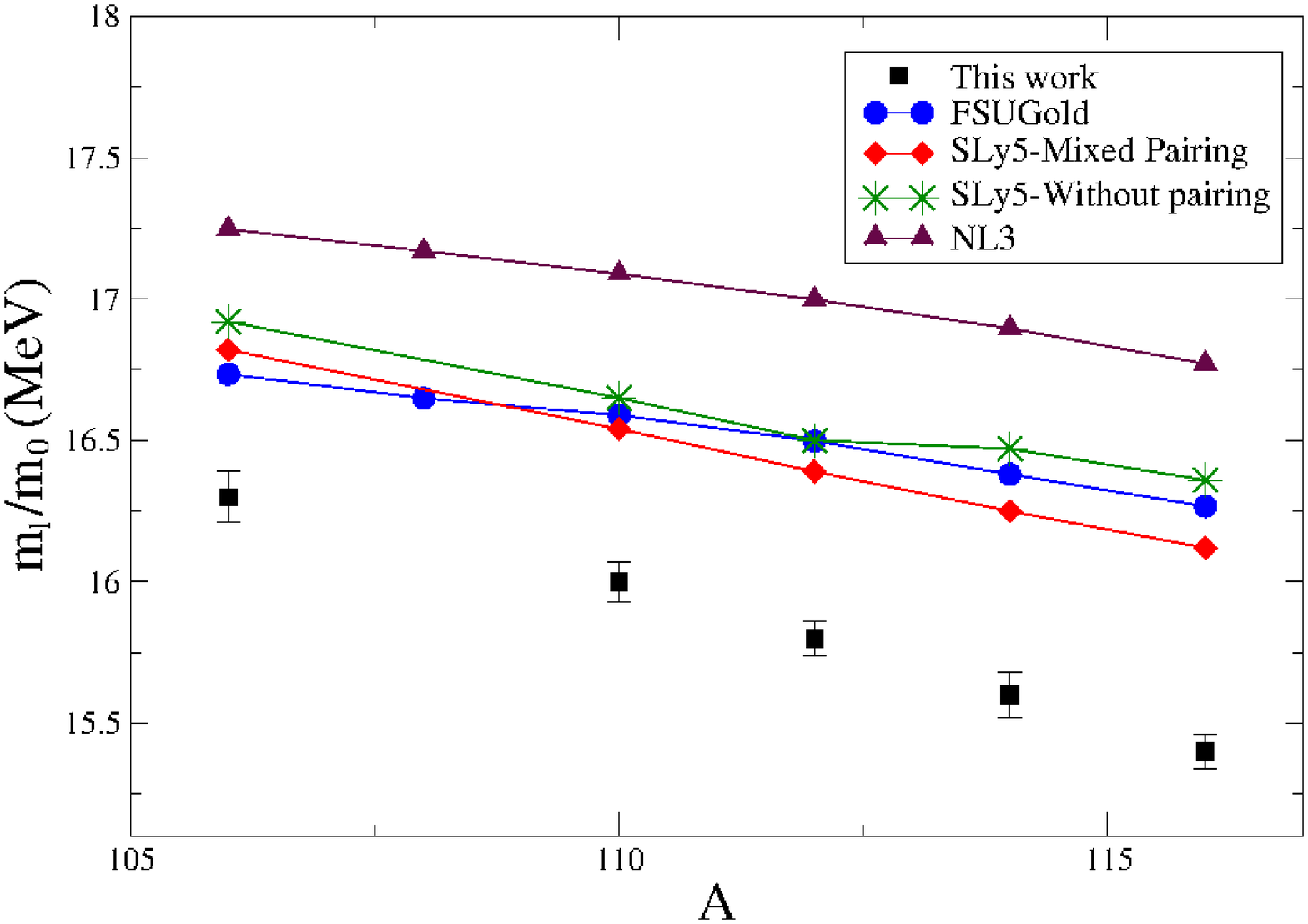}
\caption{(Color online) Systematics of the moment ratio, $m_1/m_0$ for the ISGMR
strength distributions in the Cd isotopes investigated in this work. The
experimental results (squares) are compared with
relativistic calculations performed using the FSUGold (circles) and NL3
interaction(triangles). Also presented are results from
non-relativistic calculations performed using the SLy5 parameter set in the
HFB+QRPA formalism with and without the mixed pairing interaction (diamonds and
stars, respectively) \cite{colo2}.
The solid lines are to guide the eye. \label{gl}}
\end{figure}

\noindent
centroid energies in ${}^{90}$Zr, ${}^{144}$Sm, and ${}^{208}$Pb \cite{Piekarewicz:2009gb}. Yet, this same model significantly overestimates the ISGMR energies in the Sn isotopes \cite{Piekare2007}.  Perhaps not surprisingly, our present theoretical results overestimate the centroid energies in the nearby Cd isotopes as well. The use of the NL3 effective interaction, with an incompressibility coefficient significantly larger than FSUGold, exacerbates the discrepancy between theory and experiment even further. Likewise, in a recently available calculation \cite{colo2} within the Skyrme Hartree-Fock+BCS and quasiparticle RPA with the SLy5 parameter set ($K_\infty$=230 MeV, which, incidentally, reproduces the ISGMR in $^{208}$Pb very well), the centroids of ISGMR strength distributions in the Cd isotopes (also shown in Fig. 3) are, again, significantly larger than the experimentally-obtained results. Thus, the question originally posed in Refs.\,\cite{garg2007,Piekare2007,Piekarewicz:2009gb} of {\sl ``Why are the Sn isotopes so Fluffy''} extends to the cadmium isotopes as well.

We conclude this theoretical discussion with a brief comment on the possible role of superfluid (or pairing) correlations on the softening of the isoscalar monopole response in the Sn and Cd isotopes.  This was investigated previously by Civitarese et al. \cite{Civ1991} and more recently by Li et al. \cite{Jun2008} and Cao {\em et al.} \cite{colo2}. Indeed, even with inclusion of the pairing effects, using a mixed pairing interaction \cite{colo2}, the centroids of the ISGMR remain well above the experimental values (see Fig.~3)--the net effect appears to be that of lowering the centroid by only $\sim$100 keV in $^{106}$Cd to a maximum of $\sim$240 keV in $^{116}$Cd. Thus, the impact of superfluid correlations on the compressibility of a fermionic droplet remains an interesting open question~\cite{Khan2009} to date, and, in spite of significant theoretical effort \,\cite{Jun2008,Avdeenkov:2008bi,Khan2009,Khan:2010mv, julich,Vesely:2012dw,colo2}, no single approach has been able to simultaneously describe the centroid energies in ${}^{90}$Zr, ${}^{208}$Pb, and the Sn/Cd isotopes. The remaining challenge, therefore, is not only to describe the distribution of monopole strength along the isotopic chain in tin and cadmium, but to do so without sacrificing the enormous success already achieved in reproducing a host of ground-state observables and collective modes.

\begin{figure}
\vspace*{0.5cm}
\includegraphics[width=0.48\textwidth]{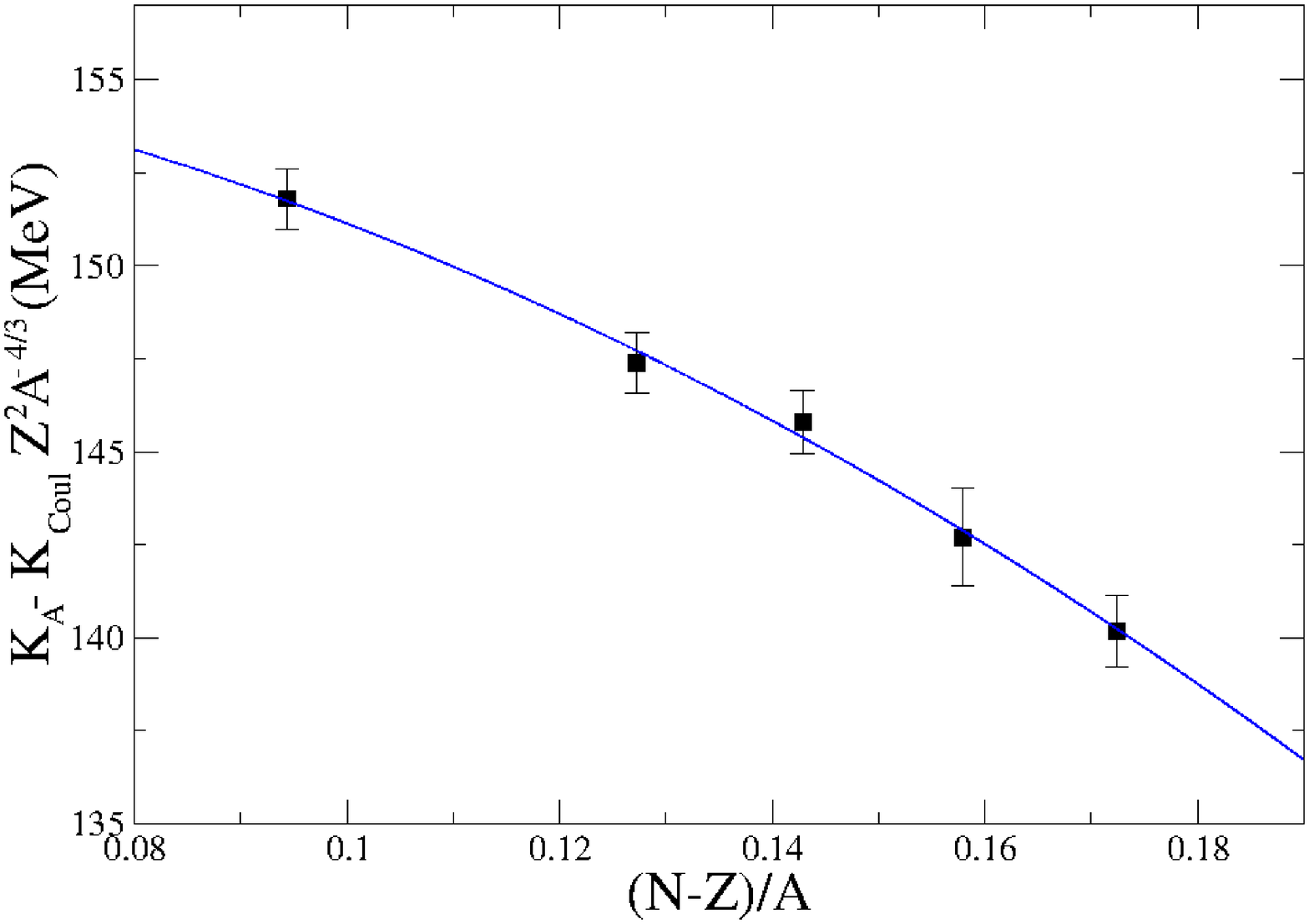}
\caption{(Color online) The difference $K_A-K_{Coul}Z^{2}A^{-4/3}$ in the
Cd isotopes investigated in this work
plotted as a function of the asymmetry parameter, $(N-Z)/A$. The values of
$K_A$ have been derived using the customary moment ratio $\sqrt{m_1/m_{-1}}$ for
the energy of ISGMR, and a value of 5.2$\pm$0.7 MeV has been used for $K_{Coul}$
\cite{Sagawa2007}. The solid line represents a quadratic fit to the data.\label{gl2}}
\end{figure}

As pointed out earlier, the asymmetry term in the nuclear incompressibility can be studied using Eq.~2. It should be noted that the first term varies very little over the isotopic chains of Sn and Cd for which the neutron-proton asymmetry changes by $81\%$ and $83\%$ respectively, and that the $K_{Coul}$ is essentially model independent. Therefore, an approximately quadratic relation
between [$K_A - K_{Coul}Z^{2}A^{-4/3}$] and the neutron-proton asymmetry,
$(N-Z)/A$, can
be applied to fit the experimental data. The result of such a quadratic fit is
shown in Fig.~4. The asymmetry term in the nuclear incompressibility, $K_\tau$,
obtained from this fit is  $-555 \pm$75 MeV; the quoted error includes the
effect ($\sim$20 MeV) of the uncertainty in the $K_{Coul}$ term.
This result confirms, and is in excellent agreement with, the value $K_\tau=-550 \pm$100 MeV obtained from the Sn isotopes \cite{Tao2007,Tao2010}. This value is also consistent with the $K_\tau=-370 \pm$120 MeV obtained from the analysis of the isotopic transport ratios in medium-energy heavy-ion reactions \cite{Chen2009}, $K_\tau=-500^{+120}_{-100}$ MeV obtained from constraints placed by neutron-skin data from anti-protonic atoms across the mass table \cite{Centelles2009}, and $K_\tau=-500 \pm$50 MeV obtained from theoretical calculations using different Skyrme interactions and relativistic mean-field (RMF) Lagrangians \cite{Sagawa2007}.

In summary, we have measured the strength distributions of the isoscalar giant monopole resonance in the even-A $^{106, 110-116}$Cd isotopes via inelastic scattering of $\alpha$ particles at extremely-forward angles, including 0$^{\circ}$, with the aim to put on further experimental footing the results obtained earlier for the ISGMR in the Sn isotopes. The centroids of the ISGMR have been calculated in the relativistic RPA using the NL3 and FSUGold effective interactions. The calculated centroids for the Cd isotopes are significantly larger than the experimentally obtained values, similar to the results obtained for the Sn isotopes, leaving the question of the ``softness'' of the open-shell nuclei open still. The value of $-555 \pm$75 MeV for the asymmetry term in the nuclear-matter incompressibility, $K_\tau$, extracted from this measurement confirms the value obtained from the study of the Sn isotopes.

This work has been supported in part by the National Science Foundation (Grant Nos. PHY07-58100 and PHY-1068192 ) and by the Department of Energy under contact no. DE-FD05-92ER40750.

\vspace*{1.5cm}


\begin{thebibliography}{99}
\bibitem{Prakash2001} J. M. Lattimer and M. Prakash, Astrophys. J. {\bf 550} (2001) 426
\bibitem{HoroPiekare2001} C. J. Horowitz and J. Piekarewicz, Phys. Rev. Lett. {\bf 86} (2001) 5647
\bibitem{BohrMott1975} A. Bohr and M. Mottelson, Nuclear Structure Vol. II (Benjamin, New York, 1975)
\bibitem{Strin82} S. Stringari, Phys. Lett. B {\bf 108} (1982) 232
\bibitem{trein81} J. Treiner et al., Nucl. Phys. A {\bf 371} (1981) 253
\bibitem{Blazoit80} J. P. Blaizot, Phys. Rep. {\bf 64} (1980) 171
\bibitem{Patra2002} S. K. Patra et al., Phys. Rev. C {\bf 65} (2002) 044304
\bibitem{Sagawa2007} H. Sagawa et al., Phys. Rev. C {\bf 76} (2007) 034327
\bibitem{cente09} J. Piekarewicz and M. Centelles, Phys. Rev. C {\bf 79} (2009) 054311
\bibitem{sharma} M. M. Sharma et al., Phys. Rev. C {\bf 38} (1988) 2562
\bibitem{Tao2007} T. Li et al., Phys. Rev. Lett. {\bf 99} (2007) 162503
\bibitem{Tao2010} T. Li et al., Phys. Rev. C {\bf 81} (2010) 034309
\bibitem{Chen2009} Lie-Wen Chen et al., Phys. Rev. C {\bf 80} (2009) 014322
\bibitem{Centelles2009} M. Centelles et al., Phys. Rev. Lett. {\bf 102} (2009) 122502
\bibitem{Piekare2007} J. Piekarewicz, Phys. Rev. C {\bf 76} (2007) 031301
\bibitem{garg2007} U. Garg et al., Nucl. Phys. A {\bf 788} (2007) 36c
\bibitem{prelim1} D. Patel et al., Bull. Am. Phys. Soc. {\bf 55} (2010) DNP.CH7
\bibitem{prelim2} U. Garg, Acta Phys. Pol. B {\bf 42} (2011) 659
\bibitem{Fujiwara99} M. Fujiwara et al., Nucl. Instrum. Meth. Phys. Res. A {\bf 422} (1999) 484
\bibitem{Fujita99} H. Fujita et al., Nucl. Instrum. Meth. Phys. Res. A {\bf 469} (2001) 55
\bibitem{Itoh2003} M. Itoh et al., Phys. Rev. C {\bf 68} (2003) 064602
\bibitem{xxxx} G. R. Satchler and D. T. Khoa, Phys. Rev. C {\bf 55} (1997) 285
\bibitem{Nayak2006} B. K. Nayak et al., Phys. Lett. B {\bf 637} (2006) 43
\bibitem{Youngblood} Y.-W. Lui et al., Phys. Rev. C {\bf 69} (2004) 034611
\bibitem{RickDOLF} L. D. Rickerston, The folding program DOLFIN (1976) ({\em unpublished})
\bibitem{Brown80} M. Rhoades-Brown et al., Phys. Rev. C {\bf 21} (1980) 2417
\bibitem{nndc} http://www.nndc.bnl.gov
\bibitem{HarakehBook} M. N. Harakeh and A. van der Woude, Giant Resonances Fundamental High-Frequency Modes of Nuclear Excitations (Oxford University Press, New York, 2001)
\bibitem{Satchler87} G. R. Satchler, Nucl. Phys. A {\bf 472} (1987) 215
\bibitem{Bonin84} B. Bonin et al., Nucl. Phys. A {\bf 430} (1984) 349
\bibitem{Uchida2004} M. Uchida et al., Phys. Rev. C {\bf 69} (2004) 051301 (R)
\bibitem{Data} S. S. Dietrich and B. L. Berman, At. Data Nucl. Data Tables {\bf 38} (1988) 199
\bibitem{bran} S. Brandenburg et al., Nucl. Phys. A {\bf 466} (1987) 29
\bibitem{matyas1} M. Hunyadi et al., Phys. Lett. B {\bf 576} (2003) 253
\bibitem{Nayak2009} B. K. Nayak et al., Phys. Lett. B {\bf 674} (2009) 281
\bibitem{colo2} Li-Gang Cao, H.Sagawa, and G. Col\`{o}, arXiv:1206.6552 (2012);
G. Col\`{o}, {\em private communication}.
\bibitem{Lalazissis:1996rd} G.~A. Lalazissis, J. Konig, and P. Ring, Phys. Rev. C {\bf 55} (1997) 540
\bibitem{Todd-Rutel:2005fa} B.~G. Todd-Rutel and J. Piekarewicz, Phys. Rev. Lett. {\bf 95} (2005) 122501
\bibitem{Piekarewicz:2001nm} J. Piekarewicz, Phys. Rev. C {\bf 64} (2001) 024307
\bibitem{Piekarewicz:2009gb} J. Piekarewicz, J. Phys. G {\bf 37} (2010) 064038
\bibitem{Civ1991} O. Civitarese et al., Phys. Rev. C {\bf 43} (1991) 2622
\bibitem{Jun2008} Jun Li et al., Phys. Rev. C {\bf 78} (2008) 064304
\bibitem{Khan2009} E. Khan, Phys. Rev. C {\bf 80} (2009) 057302
\bibitem{Avdeenkov:2008bi} A. Avdeenkov et al., arXiv:0808.0478 (2008)
\bibitem{Khan:2010mv} E. Khan et al., Phys. Rev. {\bf C 82} (2010) 024322
\bibitem{julich} V. Tselyaev et al., Phys. Rev. C {\bf 79} (2009) 034309
\bibitem{Vesely:2012dw} P. Vesle\'{y} et al., Phys. Rev. C {\bf 86} (2012) 024303

\end{thebibliography}
\end{document}